\documentclass[epj]{svjour}
\usepackage{amsmath,amssymb,bm}
\usepackage{graphics}
\usepackage{graphicx}
\usepackage{tabularx}
\usepackage{multirow}
\usepackage{psfrag}
\usepackage{comment}
\usepackage{pifont}

\usepackage{url}
\usepackage{epsfig}
\usepackage{array}
\usepackage{epstopdf}
\usepackage{hyperref}
\usepackage[english]{babel}

\usepackage{epsfig}
\usepackage{color}

\begin{document}
%
%
\title{Merits and constraints of low-${\bm K^2}$ experimental data for the proton radius determination}

\author{
M.~Hoballah\inst{1}, 
S.~Cholak\inst{1,2}, 
R.~Kunne\inst{1}, 
C.~Le~Galliard\inst{1}, 
D.~Marchand\inst{1}, 
G.~Qu\'em\'ener\inst{3}, 
E.~Voutier\inst{1}, 
J.~van~de~Wiele\inst{1} 
}
\institute{
Institut de Physique Nucl\'eaire, Universit\'es Paris-Sud \& Paris-Saclay, CNRS/IN2P3, 91406 Orsay, France \and
Physics Faculty, Taras Shevchenko National University of Kyiv, Kyiv, Ukraine, 01601 \and
Normandie Univ, ENSICAEN, UNICAEN, CNRS/IN2P3, LPC Caen, 14000 Caen, France
}

\date{Received: date / Revised version: date} 

\abstract{ %
The question of the determination of the proton charge radius $R_p$ from electron scattering data led to an unprecedented 
experimental effort for measurements of the electric form factor of the proton at low and very low momentum transfer in 
electron and muon elastic scattering. On the basis of basic properties of densities and fitting bias considerations, a 
procedure is developed in order to evaluate the impact of forthcoming data on $R_p$. Particularly, it is shown that a 0.1\% 
precision on these future cross section data is necessary to establish indisputably the $R_p$-value as determined from lepton 
scattering. The ProRad (Proton Radius) experiment at the PRAE (Platform for Research and Applications with Electrons) facility 
in Orsay is further discussed, especially the experimental method to meet this stringent constraint. 
} 

\authorrunning{M.~Hoballah {\it et al.}}
\titlerunning{Proton radius puzzle}

\maketitle

%
%
\hyphenation{me-thod}
\hyphenation{wi-thin}
\hyphenation{ca-lo-ri-me-ter}
\hyphenation{char-ge}

%
%
\section{Introduction}

The proton charge radius puzzle~\cite{Poh14} arose from the significant disagreement between measurements of the proton radius from the Lamb shift of muonic hydrogen~\cite{Poh10} and from electron scattering experiments~\cite{Ber10} as well as the spectroscopy of atomic hydrogen~\cite{Moh12}. The spectroscopy of muonic hydrogen by the CREMA Collaboration~\cite{Poh10} effectively turned upside down the world of atomic and subatomic physics, not only because of the measured value of the proton charge radius but even more so because of the superb quality of the experimental result, which does not leave room for any ambiguity. This result was confirmed by a further measurement of the CREMA Collaboration from the spectroscopy of muonic deuterium~\cite{Poh16}. However, the recent release of two high accuracy atomic hydrogen measurements, one~\cite{Bey17} in better agreement with the CREMA value and the other~\cite{Fle17} with the CODATA value~\cite{Moh16}, is further questioning the spectroscopy measurements of the proton radius.

As a consequence of this puzzle, the electron elastic scattering technique was deeply revisited, from its basic principles up 
to its interpretation. In contrast to atomic spectroscopy, lepton scattering ($e^{\pm}$ or $\mu^{\pm}$) off protons provides 
an indirect measurement of the proton radius obtained from the slope of the electric form factor at zero-momentum transfer 
($K^2$,  expressed in fm$^{-2}$ units)\footnote{In this work, we use bold small-letters ($\bm{v}$) for three-vectors, small 
letters ($v = \vert \bm{v} \vert$) for the module of three-vectors, and capital letters ($V \equiv (v_0, \bm{v} )$) for 
four-vectors.}. Within a relativistic approach, the quantity measured by scattering experiments can be formally related to 
the char-ge radius of the proton in the proton Breit frame where the zero-component of the electromagnetic current involves 
solely the electric form factor $G_E(-K^2)$. In this frame, $K^2$=$-k^2$ and $G_E(k^2)$ is the exact Wigner transform of the 
quantum charge distribution. Whether the electron scattering radius is exactly the same as the spectroscopy radius can 
however be questioned. It has been suggested for a long time that some relativistic effects~\cite{Lic70} occur and may affect 
data interpretation~\cite{Kel02}. The internal structure of the nucleon further complicates the problem. The light-cone 
framework offers a model independent interpretation of form factors in terms of transverse densities~\cite{Mil07}, but does 
not provide a clear link (if any) with the spectroscopy radius. These relativistic issues are still not satisfactorily solved 
and are related in part to the fundamental problem of the non-relativistic reduction of a relativistic quantum field 
theory~\cite{Lor18}. \newline
Considering the existing data set, the required zero-extra-polation plays a significant role in data interpretation, as 
observed by several groups~\cite{Hil10,Kra14,Lee15}. The $K^2$-domain of interest for data interpolation is a further concern, 
especially because of the sensitivity of the obtained proton radius to the considered maximum  $K^2$~\cite{Hig16,Gri16}. 
Ideally, one would like to obtain a result independent of the functional form of the  extrapolation and of the data momentum 
region. A pseudo-data method was recently proposed to efficiently control the sensitivity to the functional 
form~\cite{Yan18,Hay18}. A similar method is developed here to use the sensitivity to the data interpolation region in order 
to obtain precision constraints on forthcoming electron scattering experiments at low $K^2$. 

As of today, the proton charge radius puzzle remains unsolved and calls for new spectroscopy and scattering data. An unprecedented world-wide experimental effort with respect to the scattering technique is under progress, investigating on the one hand lepton non-universality and on the other hand the precision frontier. The low-$K^2$ experimental effort, under consideration in the present work, pertains to the latter and gathers several technically different projects at different facilities. The main purpose of the study presented here is to evaluate the impact of low-$K^2$ experiments on the proton radius extraction, or conversely, the constraints on upcoming experimental data in order to firmly and unambiguously establish the value of the proton radius as measured by the electron scattering technique. \newline 
The next section revisits the basics of the relationship between the form factor and the probability density function, and 
elaborates the constraints on the data intrinsic to the density functional. The following section explores the constraints on 
the data fitting procedure from higher order density moments. The previous features are combined in the remaining sections to 
define a method for evaluating the impact of low-$K^2$ experiments. Finally, the ProRad (Proton Radius) experiment under 
development at the PRAE (Platform for Research and Applications with Electrons) facility in Orsay~\cite{Mar17} is discussed, 
in particular the experimental technique allowing to reach the required precision constraint.

%
%
\section{Density and form factor}

The probability charge density function of a quantum static object of charge 1 (in unit of the electron charge) is normalized to 1 over the whole configuration space following 
\begin{equation}\label{sect10eq01}
\int d^3\bm{r} \, \rho (\bm{r}) = 1 \, .
\end{equation}
The radius $R$ attached to this density is defined as the mean value of the squared-position operator $\langle r^2 \rangle$  
\begin{equation}\label{sect10eq02}
R^2 = \, <r^2> \, = 4 \, \pi \int dr \, \rho (r) \,  r ^4 \, ,
\end{equation}
assuming a spherically symmetric density. In the non-relativistic approach, the form factor corresponding to this extended object can be defined as the Fourier transform of the spatial probability density 
\begin{equation}
G(\bm{k}) = \int d^3\bm{r} \, \rho(\bm{r}) \, \exp(- i\, \bm{k} \cdot \bm{r})
\label{eq:G}
\end{equation}
which inversion gives
\begin{equation}
\rho(\bm{r}) = \frac{1}{(2 \pi)^3} \int  d^3\bm{k} \, G(\bm{k}) \, \exp(i\, \bm{k} \cdot \bm{r} )
\label{eq:rho}
\end{equation}
with $G(0)$=1 following the normalization given in Eq.~\ref{sect10eq01}. Determining $G(\bm{k})$ from $\rho(\bm{r})$ (or the inverse) requires the knowledge over the full spatial (momentum) range of the integral. Since the experimental knowledge is limited, models have to be used to follow the procedure underlaid in Eq.~\ref{eq:G}-\ref{eq:rho}. Considering the multipole expansion of the exponential function and the series representation of the 0$^{\mathrm{th}}$-order spherical Bessel function, Eq.~\ref{eq:G} becomes
\begin{equation} \label{sect20eq08}
G(k) =  1 + \sum _{i=1}^{\infty} C_i  k^{2i} \equiv G(k^2)
\end{equation}
where
\begin{equation}
C_i = \frac{(-1)^i}{(2i + 1)!} \langle r^{2i} \rangle \label{eq:Cn}
\end{equation}
\noindent
with
\begin{equation}
\langle r^{2i} \rangle = 4\pi \int dr \, \rho(r) \, r^{2i+2} \, .\label{eq:rms}
\end{equation}
This establishes that the form factor depends on the even moments of the spatial probability density, and that it can be  expressed as a function of $k^2$. It should be stressed that the $k^2$-dependence results directly from using the series  expansion of the Bessel function. Consequently, it is limited to the validity domain of this representation. The derivative of Eq.~\ref{sect20eq08} with respect to $k^2$ writes 
\begin{equation}\label{derivative}
\frac{d G(k^2)}{d k^2} =  \sum _{i=1}^{\infty} i C_i k^{2(i-1)}
\end{equation}
from which the radius is deduced as
\begin{equation}
R = \sqrt{\langle r^{2} \rangle} = \sqrt{- \, 6 \, \left. \frac{d G(k^2)}{d k^2} \right\vert_{k^2=0} } \, .
\label{eq:rmsinexpansion}
\end{equation}
Hence, the radius is commonly defined as the slope of the form factor at $k^2$=$0$. It corresponds in that sense to the slope 
of a mathematical function deduced from experimental data. When the zero $k^2$-point is experimentally inaccessible, as in the 
case of elastic electron  scattering, the determination of the radius requires an extrapolation of the knowledge acquired in 
domains where $k^2 \neq 0$. Minimizing systematical effects, the measurement of the radius through extrapolation from a low 
$k^2$ domain close to the physical point $k^2$=$0$ is thought to be robust. This motivates the current world wide experimental 
effort to measure the proton electric form factor at low momentum transfer.

The previous description defined the experimental method applied in elastic electron scattering experiments to determine the radius of nucleons and nuclei. While established on firm theoretical grounds, the experimental realization of this approach necessarily suffers limitations, in particular the sensitivity to the data fitting procedure~\cite{Lee15,Hig16,Gri16}. One can understand this feature by rewriting Eq.~\ref{sect20eq08} as 
\begin{equation}
G(k^2) = 1 + C_1 k^2 \left( 1 +\sum_{i>1}^{\infty} \frac{C_i}{C_1} k^{2(i-1)} \right) \, , \label{eq:expfactorised}
\end{equation}
where the sum in the right-hand side represents the relative correction to the first order term $C_1$, originating from higher order terms in the expansion. The fitting of experimental data is nothing else than a mathematical reproduction of Eq.~\ref{eq:expfactorised}. Since the correction term scales with powers of $k^2$, it is intuitive that the higher $k^2$, the larger the influence of the correction. In other words, extracting the radius by extrapolating the knowledge of the form factor from domains at large $k^2$ amplifies the sensitivity of the procedure to the functional form of $G(k^2)$ through the dependence on higher order terms in the series expansion. 

\begin{figure}[t!]
\centering
\includegraphics[width=0.995\columnwidth]{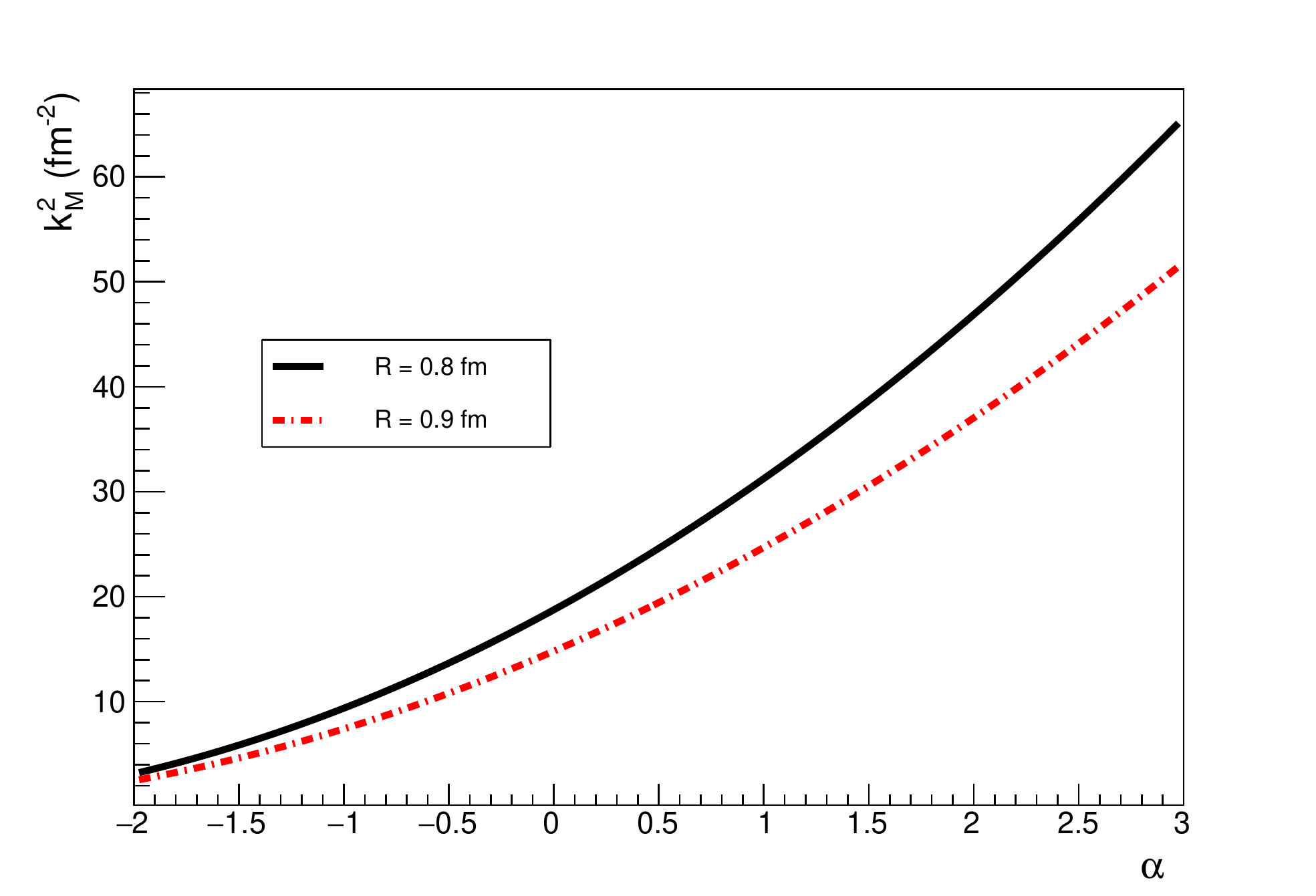}
\caption{Density model sensitivity of the momentum boundary resulting from the convergence criterion (Eq.~\ref{kalfa}).}
\label{kMAX}
\end{figure}
The magnitude of the correction terms in Eq.~\ref{eq:expfactorised} involves the ratio of the moments of the spatial 
probability density, and as such is model dependent. Considering the limited $k^2$ experimental knowledge, the choice of a 
density model $\rho(r)$ is a delicate issue. However, any model should represent a physically possible case, particularly, 
it should ensure the convergence of the $G(k^2)$ expansion at high order values, that is 
\begin{equation}\label{sect1}
\lim_{i \to \infty} k^2 \left\vert \frac{C_{i+1}}{C_{i}} \right\vert < 1 \, .
\end{equation}
For a given density model, the $C_i$'s coefficients can be evaluated to any order in the expansion. Considering for instance the generic form 
\begin{equation}
\rho(r) =  \rho_0\,r^\alpha \exp(-\lambda r) \label{eq:rhorxample}
\end{equation}
where $\rho_0$ is a normalization constant defined as 
\begin{equation}
\rho_0 = \frac{1}{4 \pi} \, \frac{\lambda^{3+\alpha}}{\Gamma(3+\alpha)} \, , \label{eq:rhorxampleform}
\end{equation}
the moments of the spatial density write
\begin{equation}
\langle r^{2i} \rangle = \frac{1}{\lambda^{2i}} \, \frac{\Gamma(2i+3+\alpha)}{\Gamma(3+\alpha)} \, , \label{eq:rmsexact}
\end{equation}
and specifically, the radius is given by
\begin{eqnarray}\label{sect202eq05}
R = \sqrt {\frac{(3+\alpha ) (4+\alpha )}{\lambda ^2}} \, .
\end{eqnarray}
The convergence criterion of the series for the density model and radius considered here writes
\begin{equation}\label{kalfa}
k^2 < k^2_M = \frac{(3+\alpha ) (4+\alpha )}{R^2}
\end{equation}
which limits the $k^2$-range over which the series representation of the Fourier transform can be used. Since data fitting mimics the series representation, the convergence criterion effectively restricts the ability to extract any possible density functionals from any measured  $k^2$-domain. This is expressed in Fig.~\ref{kMAX} which shows the $\alpha$-evolution of $k^2_M$ for a radius variable in the range 0.8-0.9~fm. At a fixed $k^2_M$, the relation Eq.~\ref{kalfa} acts as the minimum acceptable $\alpha$-parameter for data fitting. The small $k^2_M$ domain only allows for the largest $\alpha$-phase-space. \newline
The restriction of the $k^2$-domain can in principle be released if the form factor data are fitted against the Fourier transform of the density. Considering as example Eq.~\ref{eq:rhorxample}, the form factor can be derived analytically from the Fourier transform of the density as
\begin{equation}
G(k) = \frac{\lambda^{3+\alpha}}{{(k^2+\lambda^2)}^{1+\alpha/2}} \, \frac{\sin \left[ (2+\alpha) \arctan{\left( \frac{k}{\lambda} \right)} \right] }{(2+\alpha) \, k} \label{eq:ftrans}
\end{equation}
which indicates that $G(k)$ is most generally a $k$-dependent function. Apart from the specific case of the dipole form factors ($\alpha$=0), it  is only in the limit $k \ll \lambda$ that Eq.~\ref{eq:ftrans} reduces to a pure $k^2$-dependence, consistently with the validity range of the serial expansion. However, such an approach remains strongly model dependent since the true shape of the density is unknown. 

\begin{figure}[t!]
\centering
\includegraphics[width=0.995\columnwidth]{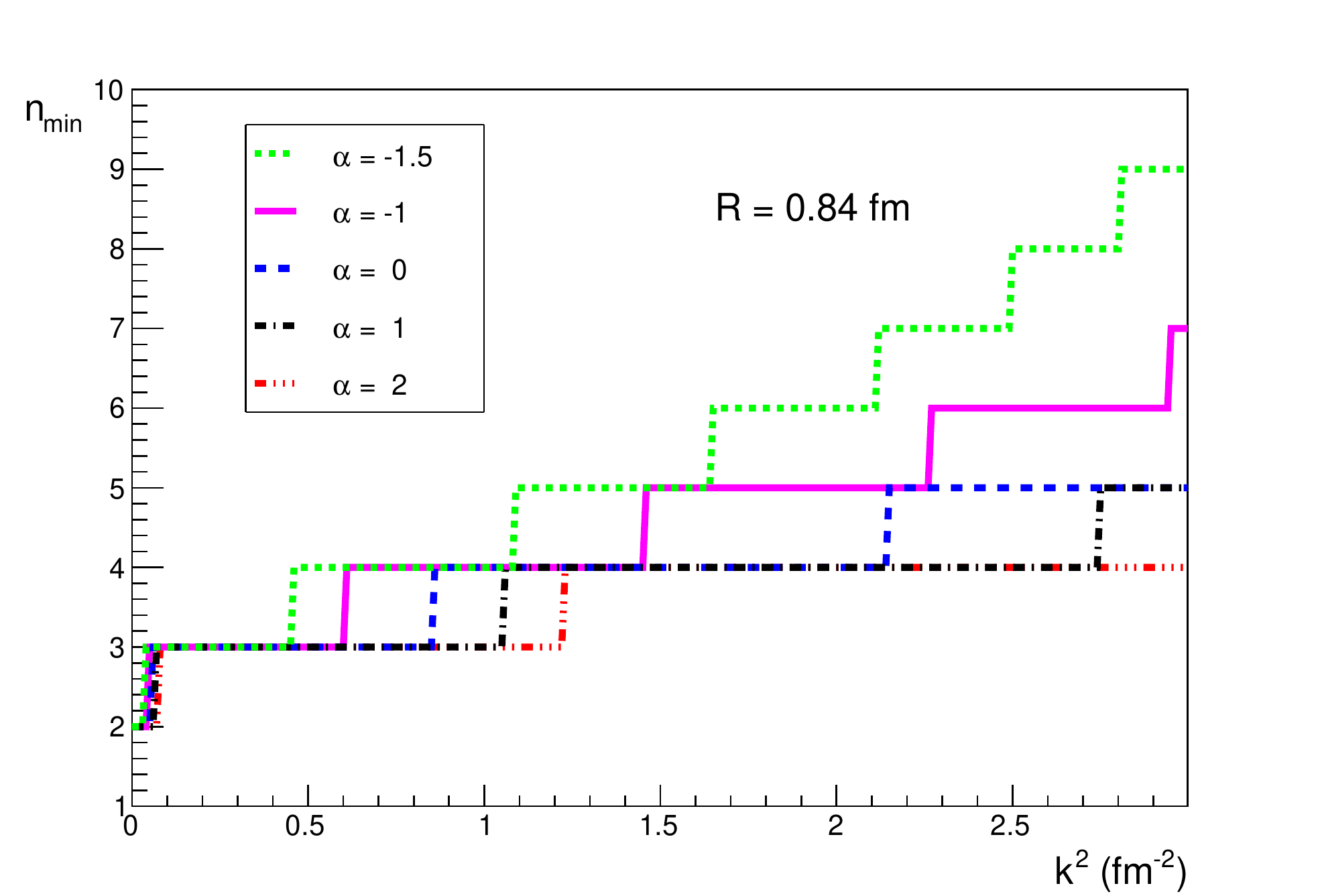}
\caption{The minimum expansion order $n_{min}$ required to describe $G(k^2)$ with a 5$\times$10$^{-4}$ accuracy for different models  parameterized by $\alpha$~(Eq.~\ref{eq:rhorxample}), and assuming the radius value $R$=$0.84$~fm.}
\label{fig:polyCorrection}
\end{figure}

For a given density model, the relative correction from each term in Eq.~\ref{eq:expfactorised} can be evaluated. Requiring the value of the form factor at $k^2$ to be precise within some accuracy corresponds to limiting the effects of the higher order terms. This leads to a minimal order of the expansion which intuitively increases with $k^2$. Figure~\ref{fig:polyCorrection} shows the $k^2$-evolution of the $n_{min}$ minimal order required to correctly describe the form factor at $k^2$ with a 5$\times$10$^{-4}$ accuracy, that is
\begin{equation}
\frac{C_i}{C_1} \, k^{2(i-1)} \le 5 \times 10^{-4} \, .
\end{equation}
Considering a fixed radius of $R$=$0.84$~fm, the different $\alpha$-models are compared in Fig.~\ref{fig:polyCorrection}. There exists a $k^2$-domain (up to $\sim$2~fm$^{-2} \approx 4 m_{\pi}^2$) where $n_{min}$ is weakly depending on the specific density model. Such a domain is of experimental interest since it does not suffer from density related bias, {\it i.e.} this restricted $k^2$-domain is not limited by the convergence criterion and does not bias the shape of the density that can be extracted from data. Note that the 5$\times$10$^{-4}$ accuracy constraint does only affect the specific value of $n_{min}$ but not the global behaviour. Additionally, the restriction to this limited kinematical domain naturally preserves the analyticity properties of the proton form factors~\cite{Hil10}. 

These features are strong indications that the low-momentum transfer region up to 2~fm$^{-2}$ is most suitable for a robust determination of the radius through the $k^2$-dependence of the form factor.  

%
%
\section{Data fitting and radius}

The previous section established the existence of a density dependent relationship between the series representation of the form factor up to a certain momentum transfer $k^2_M$, and the expansion order required to contain higher order effects on $C_1$. $n_{min}$ should not be confused with the order of a polynomial fit. However, the data fitting procedure can be inspired from these density dependent considerations. While a minimal order limit is suggested, there is no maximum order limit that can be used to describe the data. Fits with too many parameters may still give acceptable results with excellent $\chi^2$ and confidence probability but different radius value~\cite{Lee15,Ber14}. However, using an expansion order higher than what is expected to describe the data would generate overfitting problems and could lead to biased results.

This feature is illustrated further with a pseudo-data procedure. A thousand data sets have been generated following a dipole parameterization assuming the radius value $R$=$0.84$~fm. The position in $k^2$ and the measurement errors of the pseudo-data are taken from the Mainz 2010 data set~\cite{Ber10-1} restricted to the momentum transfer range $k^2 < \vert K^2_{max} \vert$=$2$~fm$^{-2}$, corresponding to $n_{min}$=$4$ (Fig.~\ref{fig:polyCorrection}). Note that the specific 
value of this limit is not determinant for the present discussion. The form factor value at a given $k^2$ follows a normal distribution centered on the dipole expectation value with a gaussian width corresponding to the error at that same $k^2$. Each pseudo-data set is then fitted with polynomials of different order ($n \in [1,6]$)
\begin{equation}
P_n(k^2) = \sum_{i=0}^n a_i^n \, {(k^2)}^i \, ,\label{poln}
\end{equation}
to extract the $C_i^n$ ($i \in [1,4]$) parameters of the form factor expansion, and the corresponding moments of the density leading to 
\begin{equation}
\langle r^{2i}_n \rangle = (-1)^i (2i+1)! \, a_i^n \, .
\end{equation} 
For each moment, the compatibility criterion ($\Delta$) between the dipole reference value $\langle r^{2i}_{R}\rangle$ and the measured value $\langle r^{2i}_n \rangle$ of the i$^{\mathrm{th}}$-moment of the density from the fit of pseudo-data with the n$^{\mathrm{th}}$-order polynomial is defined as
\begin{equation}
\Delta = \frac{\langle r^{2i}_n \rangle - \langle r^{2i}_{R} \rangle}{\delta \langle r^{2i}_n \rangle} = \frac{\langle r^{2i}_n \rangle - \langle r^{2i}_{R} \rangle}{\sigma} \, ,
\label{eq:DeltaC1}
\end{equation}
where $\delta \langle r^{2i}_n \rangle$ is the moment error, and $\Delta$ is expressed in units of the standard deviation ($\sigma$).

\begin{figure}[t]
\centering
\includegraphics[width=0.995\columnwidth]{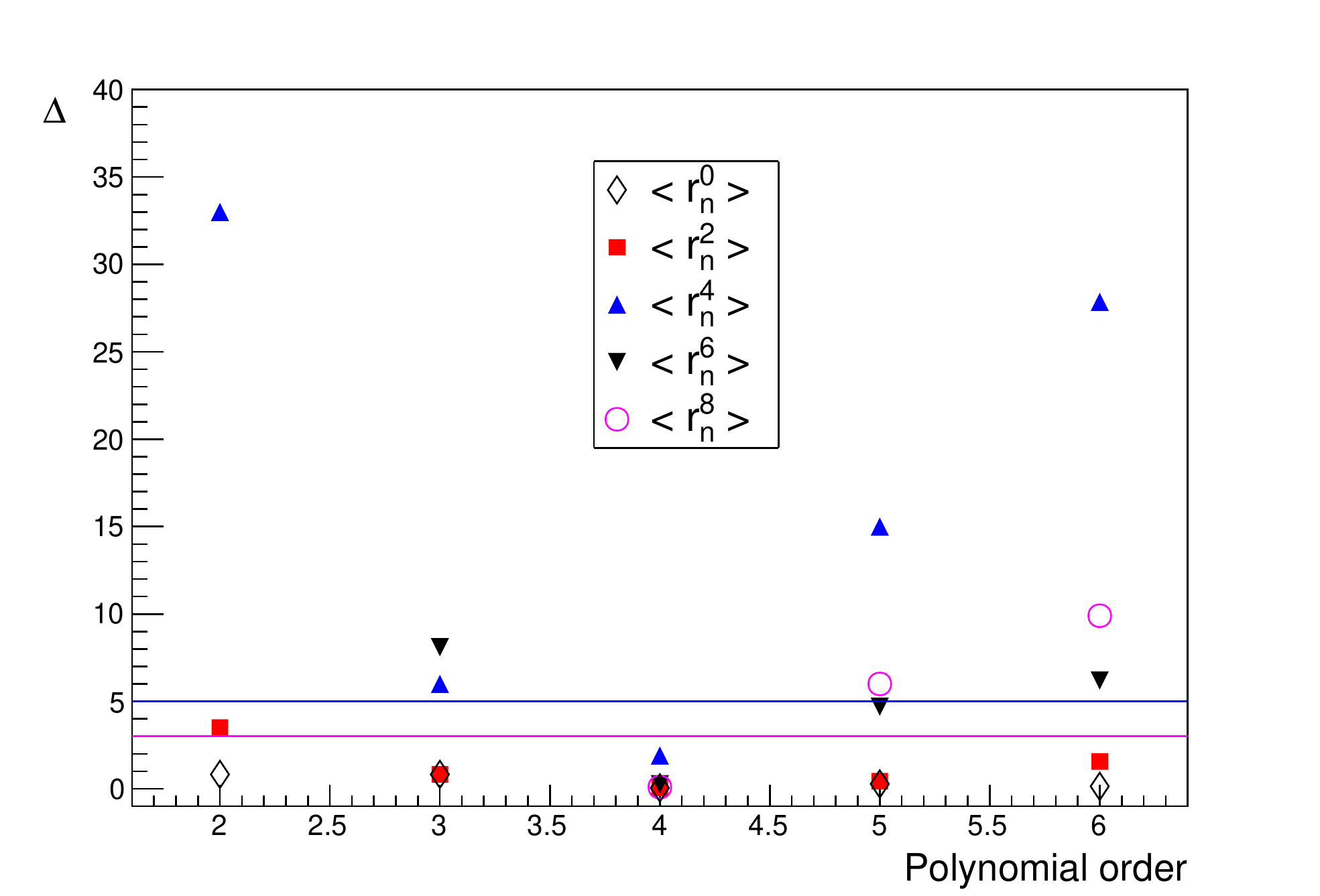}
\caption{The compatibility $\Delta$ between the fitted and the reference $\langle r^{2i}_{R} \rangle$ values as a function of the order of the polynomial fit. The lines define the 3$\sigma$ and 5$\sigma$ limits.}
\label{fig:polyCorrectionfakeData}
\end{figure}
Figure~\ref{fig:polyCorrectionfakeData} shows the compatibility criterion as a function of the order of the polynomial used to fit pseudo-data. It clearly indicates the existence of a maximum polynomial order limit $n_{max}$ at which zeroth, first and  higher order moments are recovered from the pseudo-data. Fits with a polynomial order smaller than $n_{max}$ shows systematics deviations from reference values. Fits with polynomial order higher than $n_{max}$ still recover consistent zeroth and first moment but rapidly fail higher moments. $\langle r^{2}\rangle$ recovery should be understood as the convergence of the fit toward the reference $\langle r^{2}_R\rangle$ value in the case of an infinite number of pseudo-experiments, but it does not exclude the possibility for a single experiment to find a different $\langle r^{2}\rangle$. Similarly to the first order moment, higher order moments are physics observables of the density and should be recovered by the fitting procedure. The failure of the fit for higher order moments is a direct consequence of data overfitting, {\it i.e.} fitting unphysical fluctuations from point-to-point, and indicates that a maximum polynomial order limit should be considered. The present empirical study suggests to keep $n_{max}$=$n_{min}$ to ensure a reasonable determination of the density moments.   
%
%
\section{Impact of low momentum transfer data}

\subsection{The experimental status}

The current world data set for the electric form factor of the proton with momentum transfer smaller than 2~fm$^{-2}$ 
(Tab.~\ref{tbl:dataset}) consists of 753 experimental points obtained from Rosenbluth separation 
experiments~\cite{Dud63,Bor75} or cross section measurements assuming some relationship between the electric and magnetic form 
factors ~\cite{Ber10-1,Han63,Fre66,Aki72,Mur74,Sim80,Mih17}: 22 data from early experiments prior to 
1970~\cite{Dud63,Han63,Fre66}, 43 data from experiments over the period 1970-1980~\cite{Bor75,Aki72,Mur74,Sim80}, and 688 data 
from the recent Mainz experiments~\cite{Ber10-1,Mih17}. Out of this data set, 454 data feature a momentum transfer smaller 
than 1~fm$^{-2}$. \newline Following previous sections, a fit of experimental data up to 2~fm$^{-2}$ requires a 
4$^{\mathrm{th}}$-order polynomial in $k^2$ while a 3$^{\mathrm{rd}}$-order polynomial suffices up to 1~fm$^{-2}$. The results 
of the fits of the world data with these polynomials, which generic expression is defined in Eq.~\ref{poln}, is reported in 
Tab.~\ref{tbl:fitresults2fmmainz}. Note that all the data falling in the momentum range of interest are considered with no 
specific restriction nor adjustment, as for instance the floating normalization of the Mainz data set~\cite{Ber14}. Such a 
procedure, which would allow to improve the quality of the fits and to refine the proton radius value, is not the purpose of 
the present discussion. $a_0^n$ represents the value of the form factor at zero momentum transfer, corresponding by convention 
to the unit proton electric charge and is actually reproduced with a per-mil or better accuracy by existing data. Depending on 
the maximum momentum transfer considered, the world data set supports a proton radius either in  agreement or in disagreement 
with the muonic hydrogen measurement. The objective of upcoming elastic scattering experiments in the low momentum transfer 
region is not only to allow for a new determination of the proton radius but most importantly to provide a momentum 
independent proton radius below 2~fm$^{-2}$, accurate enough to indisputably establish whether electron scattering 
measurements of the proton radius are or not consistent with the spectroscopy measurements of muonic atoms. Then, the 
potential impact of the future experiments can be quantified in terms of the precision of expected data.
\begin{table}[t!]
\setlength{\tabcolsep}{3pt}
\begin{center}
\begin{tabular}{cr||c|c}
       Data &  \multirow{2}{*}{Ref.\phantom{001}} & \multicolumn{2}{c}{Data Number} \\ 
        Set &                        & $<1$~fm$^{-2}$ & $<2$~fm$^{-2}$ \\ \hline\hline
 \ding{172} &          \cite{Ber10-1} + \cite{Mih17} & 405 & 688 \\ 
 \ding{173} &	\ding{172}\phantom{0} + \cite{Sim80} & 420 & 706 \\ 
 \ding{174} &	\ding{173}\phantom{0} + \cite{Bor75} & 423 & 713 \\ 
 \ding{175} &	\ding{174}\phantom{0} + \cite{Mur74} & 434 & 724 \\ 
 \ding{176} &	\ding{175}\phantom{0} + \cite{Aki72} & 441 & 731 \\ 
 \ding{177} &	\ding{176}\phantom{0} + \cite{Fre66} & 442 & 735 \\ 
 \ding{178} &	\ding{177}\phantom{0} + \cite{Han63} & 452 & 750 \\ 
 \ding{179} &	\ding{178}\phantom{0} + \cite{Dud63} & 454 & 753 \\ \hline\hline
\end{tabular}
\caption{The actual world data set on the electric form factor of the proton from electron scattering experiments up to $\vert K^2_{max} \vert$=1~fm$^{-2}$ or $\vert K^2_{max} \vert$=2~fm$^{-2}$. The sorting of the data combination used in Tab.~\ref{tbl:fitresults2fmmainz} follows a reverse chronological order.}
\label{tbl:dataset}
\end{center}
\end{table}
\begin{table}[t!]
\setlength{\tabcolsep}{3pt}
\begin{center}
\begin{tabular}{cc||cc|cc}
$\vert K^2_{max} \vert$ & \multirow{2}{*}{$n$} & \multirow{2}{*}{$a_0^n$} & $R_p$ & \multirow{2}{*}{$\chi^2_r$} & Data \\ 
            (fm$^{-2}$) &                      &                          &  (fm) &                             & Set \\ \hline\hline
 \multirow{8}{*}{1.0} & \multirow{8}{*}{3} & 0.99926(65) & 0.839(16) & 1.186 & \ding{172} \\ 
                      &                    & 0.99934(65) & 0.840(16) & 1.174 & \ding{173} \\ 
                      &                    & 0.99934(65) & 0.841(16) & 1.169 & \ding{174} \\ 
                      &                    & 0.99928(65) & 0.839(16) & 1.177 & \ding{175} \\ 
                      &                    & 0.99914(65) & 0.829(16) & 1.321 & \ding{176} \\ 
                      &                    & 0.99914(65) & 0.830(16) & 1.321 & \ding{177} \\ 
                      &                    & 0.99915(65) & 0.830(16) & 1.311 & \ding{178} \\ 
                      &                    & 0.99915(65) & 0.830(16) & 1.302 & \ding{179} \\ \hline\hline	      
 \multirow{8}{*}{2.0} & \multirow{8}{*}{4} & 1.00052(51) & 0.875(10) & 0.995 & \ding{172} \\ 
                      &                    & 1.00056(50) & 0.875(10) & 0.989 & \ding{173} \\ 
                      &                    & 1.00056(50) & 0.875(10) & 1.001 & \ding{174} \\ 
                      &                    & 1.00052(50) & 0.874(10) & 1.010 & \ding{175} \\ 
                      &                    & 1.00060(50) & 0.871(10) & 1.105 & \ding{176} \\ 
                      &                    & 1.00060(50) & 0.871(10) & 1.103 & \ding{177} \\ 
                      &                    & 1.00058(50) & 0.870(10) & 1.118 & \ding{178} \\ 
                      &                    & 1.00058(50) & 0.870(10) & 1.112 & \ding{179} \\ \hline\hline	      
\end{tabular}
\caption{Results of the fit of the world data about the electric form factor of the proton with a polynomial which order depends on the maximum momentum transfer considered. The last columns indicate the reduced $\chi^2$ for the corresponding data set number defined in Tab.~\ref{tbl:dataset}.}
\label{tbl:fitresults2fmmainz}
\end{center}
\end{table}

\subsection{The method}

The landscape of upcoming low momentum transfer $ep$ elastic scattering experiments is particularly rich. At the Thomas Jefferson National Accelerator Facility, the PRad experiment will provide $G_E(k^2)$ data in the momentum range  5.0$\times$10$^{-3}$-1.5~fm$^{-2}$, by measuring at small scattering angles the elastic cross section relative to the M{\o}ller cross section, using a high resolution electromagnetic calorimeter~\cite{Gas11}. The Initial State Radiation technique applied to elastic scattering at the Mainz Microtron (MAMI) facility will provide further measurements of $G_E(k^2)$ in the range 8.0$\times$10$^{-3}$-0.7~fm$^{-2}$~\cite{Mih18}. Another experiment at MAMI will use a Time Projection Chamber to detect the recoil elastic protons, investigating the momentum transfer range 2.5$\times$10$^{-2}$-0.5~fm$^{-2}$~\cite{Den18}. The Ultra Low $Q^2$ (ULQ2) project~\cite{Sud18}, under development at the Research Center for Electron Photon Science of Tohoku University, will determine $G_E(k^2)$ in the range 8.0$\times$10$^{-3}$-0.2~fm$^{-2}$ by measuring $ep$ scattering relative to $e ^{12}$C with a CH$_2$ target. At the Platform for Research and Applications with Electrons (PRAE~\cite{Mar17}) in Orsay, the ProRad experiment will measure $ep$ elastic scattering relative to M{\o}ller scattering to obtain $G_E(k^2)$ data in the range 2.5$\times$10$^{-4}$-7.5$\times$10$^{-3}$~fm$^{-2}$. While the main focus of the MUSE experiment~\cite{Gil17} at the Paul Scherrer Institut is about $\mu^{\pm} p$ elastic scattering, it will also measure $e^{\pm}p$ scattering to provide $G_E(k^2)$ data in the range 4.0$\times$10$^{-2}$-2.0~fm$^{-2}$. These experiments will constitute an impressive data set of 320 future measurements at low momentum transfer, using different experimental methods and techniques.

In order to determine the impact of these new measurements, the data set is restricted to transfer momenta below 2~fm$^{-2}$. Each planned data is characterized by its $k^2$-momentum transfer and the expected precision on the proton electric form factor $\delta G_E(k^2) / G_E(k^2)$, which is the parameter of the study. A pseudo-data set is generated according to the 3$^{\mathrm{rd}}$- or 4$^{\mathrm{th}}$-order polynomial corresponding to existing data (Tab.~\ref{tbl:fitresults2fmmainz}) leading to a reference proton radius $R_i$. Each data is further redistributed according to a gaussian whose mean and standard deviation correspond to the polynomial projection and the absolute error $\delta G_E(k^2)$, respectively. The resulting data set added to the already existing world data set is then fitted with a 3$^{\mathrm{rd}}$- or 4$^{\mathrm{th}}$-order polynomial  depending on the selected maximum momentum, which provides the measurement of the proton radius relevant to this data set. This procedure is repeated 1000 times to obtain the distribution of the measurements which gaussian adjustment provides the measured  proton radius $R_p^m$ and the error $\delta R_p^m$ attached to the initial relative error input. Finally, the comparison between $R_p^m$ and $R_i$ measures the impact of expected data. 

\subsection{Potential of upcoming data}

The merits of upcoming experiments are expressed in terms of the evaluators
\begin{eqnarray}
\Delta_1 & = & \frac{\vert R_p^m -   R_1 \vert}{\sqrt{ {\left( \delta R_p^m \right)}^2 + {\left( \delta R_1 \right)}^2 }} 
\label{eqD1} \\ 
\Delta_2 & = & \frac{\vert R_p^m -   R_2 \vert}{\sqrt{ {\left( \delta R_p^m \right)}^2 + {\left( \delta R_2 \right)}^2 }} 
\label{eqD2}
\end{eqnarray}
with
\begin{eqnarray}
R_1 & = & 0.830 \pm 0.016 \\ 
R_2 & = & 0.870 \pm 0.010 \, . 
\end{eqnarray}
\begin{figure}[t!]
\centering
\includegraphics[width=0.995\columnwidth]{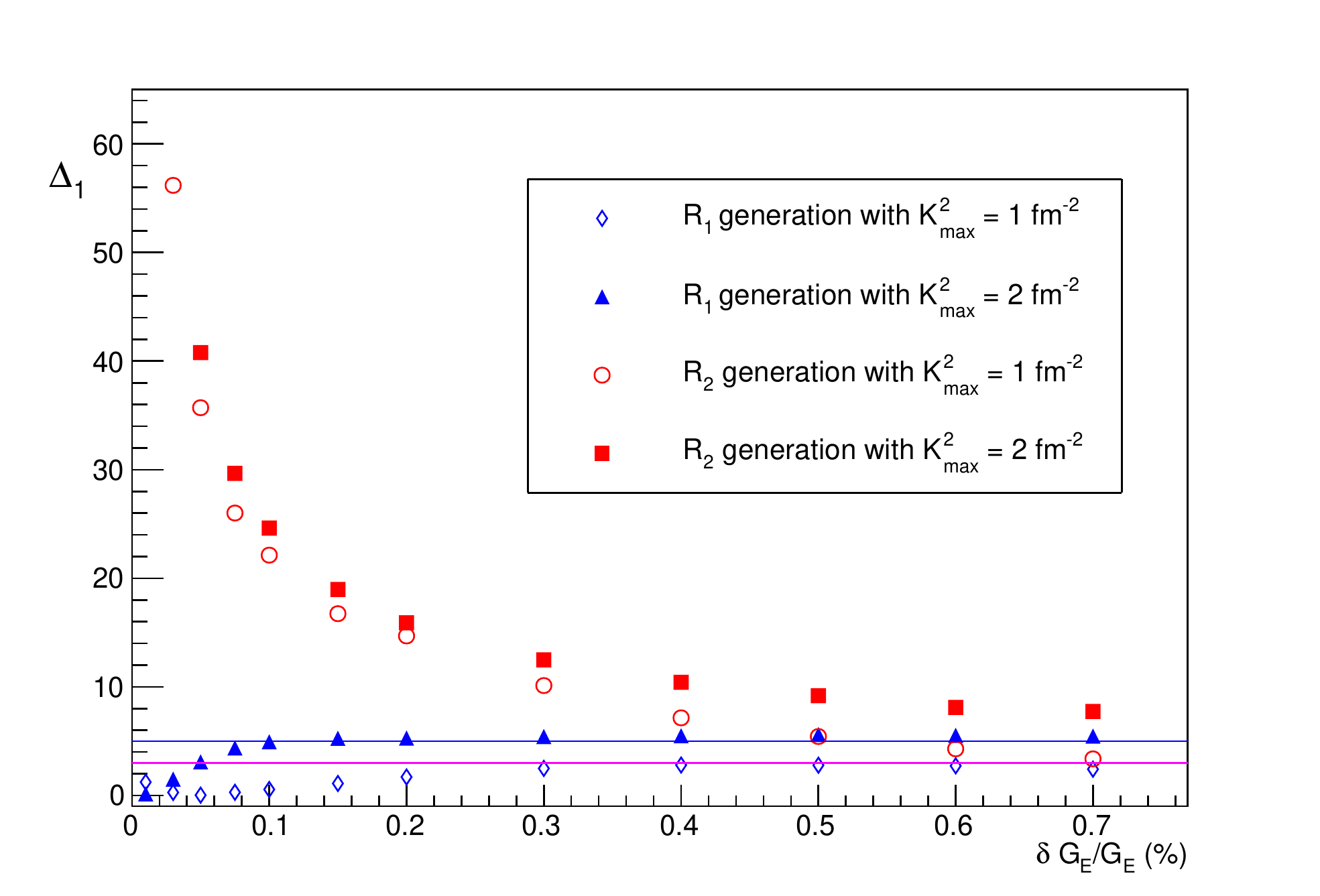}
\caption{Evolution of the $\Delta_1$ compatibility evaluator (Eq.~\ref{eqD1}) as a function of the relative accuracy of upcoming low-$k^2$ experimental data. Open and full symbols correspond to data up to 1~fm$^{-2}$ and 2~fm$^{-2}$,  respectively. Diamond and triangle symbols represent the results for pseudo-data centered on $R_1$ while circle and square  symbols stands for pseudo-data centered on $R_2$.}
\label{fig:Delta_sigma_k2max2fm}
\end{figure}
\begin{figure}[t!]
\centering
\includegraphics[width=0.995\columnwidth]{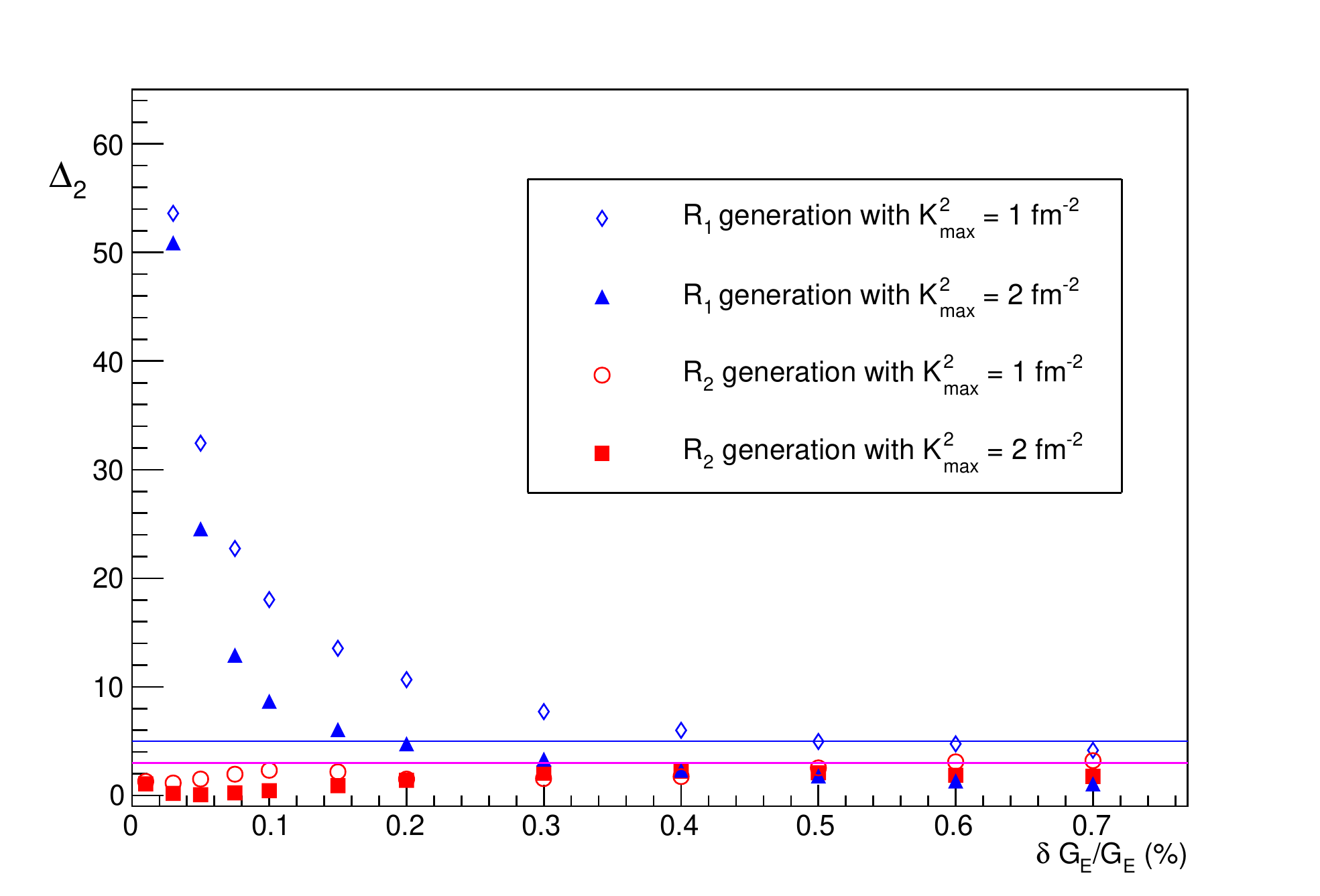}
\caption{Evolution of the $\Delta_2$ compatibility evaluator (Eq.~\ref{eqD2}) as a function of the relative accuracy of upcoming low-$k^2$ experimental data. The definition of the different symbols is the same as in Fig.~\ref{fig:Delta_sigma_k2max2fm}.}
\label{fig:Delta_sigma_k2max2fmmainz}
\end{figure}
$\Delta_1$ quantifies the compatibility of the measured proton radius, taking into account existing and expected data up to 1~fm$^{-2}$ ($R_p^1$) or 2~fm$^{-2}$ ($R_p^2$), with the $R_1$ reference value obtained from the fit of existing data up to 1~fm$^{-2}$. $\Delta_2$ quantifies the compatibility with the $R_2$ reference value obtained from the fit of existing data up to 2~fm$^{-2}$. $R_1$ and $R_2$ are chosen from Tab.~\ref{tbl:fitresults2fmmainz} considering the complete world data set. \newline
For very accurate pseudo-data, one would expect the fit procedure to converge to the radius value selected to generate 
pseudo-data. Correspondingly, one of the ($\Delta_1$,$\Delta_2$) evaluators should be small ($\le 3$) while the other should 
be large ($\ge 5$). For low accuracy, the procedure would be dominated by existing data. In  
Fig.~\ref{fig:Delta_sigma_k2max2fm} and \ref{fig:Delta_sigma_k2max2fmmainz}, the difference between open and full symbols 
indicates the effect of the maximum momentum transfer (1 or 2 fm$^{-2}$), and the difference between each open or full symbols 
indicates the radius corresponding to generated pseudo-data ($R_1$ or $R_2$). Both figures exhibit the same expected trend as 
the accuracy of pseudo-data increases: the estimator of pseudo-data generated with a value different from the reference one 
rapidly increases while the estimator of pseudo-data generated with the reference value slowly decreases. Depending on the 
generation scenario, the precision on forthcoming data required for a consistent determination of the proton radius can be 
seriously demanding: if the true radius is $R_2$, an accuracy of 0.4\% would statistically confirm this value and reject 
$R_1$, independently of $K^2_{max}$; in the opposite case, an accuracy of 0.3\% is required to confirm $R_1$ and reject $R_2$ 
when considering data up to 1~fm$^{-2}$, and as low as 0.05\% for data up to 2~fm$^{-2}$. This suggests that low-$K^2$ 
experimental data to come would conclusively establish the electron scattering measurement of the proton radius only if a 
0.1\% precision on cross section sensitive observables is reached. This is definitively a stringent constraint on experiments, 
especially considering systematical effects.

%
%
\section{The ProRad experiment at PRAE}

\subsection{The PRAE facility}

The PRAE facility~\cite{Mar17} is a multidisciplinary R\&D facility based on an electron accelerator delivering a high-performance beam (Tab.~\ref{Tab:PRAE}) with energy up to 70~MeV (Phase I) upgradable to 140~MeV (Phase II). It gathers several scientific communities involved in subatomic physics, radiobiology, instrumentation and particle accelerators for the completion of unique measurements. In the energy range 30-70~MeV, the ProRad experiment will contribute to the low-$K^2$ experimental effort by  providing high accuracy  measurements of the electric form factor of the proton in the lowest ever measured momentum transfer range. The 50-140~MeV electron energy range will allow developing pre-clinical studies of new radiotherapy methods aiming for a better treatment of cancer. Over the full energy range, PRAE beams will provide the tools to characterize and optimize instrumentation techniques for the next generation of detectors used in medical imaging, subatomic physics, particle physics, spatial technology and astrophysics.

\begin{table}[h!]
\begin{center}
\begin{tabular}{r||c|c}
   Beam parameters @ LinAc end & PRAE            & Unit   \\ \hline\hline
                Maximum Energy & 70(I) - 140(II) & MeV    \\
               Repetition rate & 50              & Hz     \\
        Relative energy spread & $< 0.2$         & \%     \\
                  Bunch charge & 0.00005-2       & nC     \\
                  Bunch length & $< 300$         & $\mu$m \\
 Number of micro-bunches/train & 1               &        \\ \hline\hline
\end{tabular}
\caption{PRAE beam performance at phase I and II.}
\label{Tab:PRAE}
\end{center}
\end{table}
The electron beam generated by a radio-frequency gun accelerates inside a 3.4~m (Phase I) long LinAc section based on high-gradient 3~GHz S-band cavities operating at a 50~Hz repetition rate~\cite{Bar17}. Two of these sections will be installed in Phase II of the project. A beam energy compression section constituted of a magnetic chicane and a dechirper structure, follows the accelerating section, and allows to reduce the beam energy dispersion to a few 0.01\%. Flexible beam optics is ensured by several magnets to cope the different beam characteristics and operation modes depending on the application. The PRAE facility is currently under construction at the Laboratoire de l'Acc\'el\'erateur Lin\'eaire in Orsay. First beam delivery is planned for 2021.

\subsection{The ProRad experimental setup}

\begin{figure}[t!]
\centering
\includegraphics[width=0.950\columnwidth]{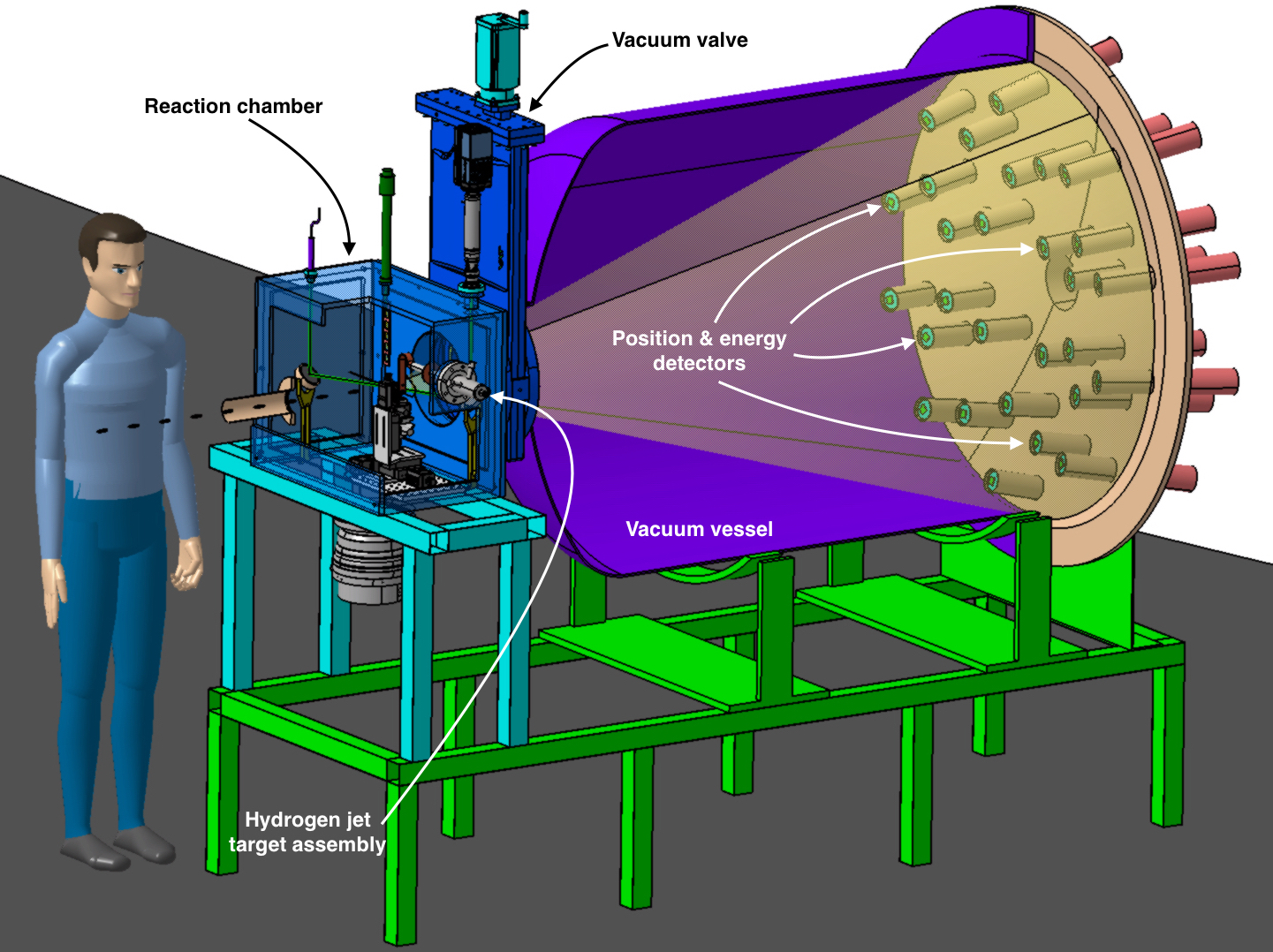}
\caption{The ProRad experimental setup.}
\label{setup}
\end{figure}

Following the accelerating and beam characterization sections of the PRAE machine, the ProRad experimental setup (Fig.~\ref{setup}) features a reaction chamber followed by a vacuum vessel closed by an end-cap supporting the detector elements. The experiment is designed to measure with a 0.1\% accuracy the ratio 
\begin{equation}
\rho_{\sigma} (K^2) = \frac{d^2 \sigma_{ep}}{d\Omega} \bigg{/} \frac{d^2 \sigma_{ee}}{d\Omega}
\end{equation}
representing the cross section of the elastic electron scattering off protons ($d^2 \sigma_{ep}/d\Omega$) relative to the 
M{\o}ller cross section ($d^2 \sigma_{ee}/d\Omega$). The reaction chamber hosts a windowless target made of a 15~$\mu$m 
diameter solid hydrogen jet~\cite{Cos12}, effectively concretizing a pure and nearly point-like target. Scattered electrons 
propagate over 2~m inside the vacuum vessel till the detection area. The ProRad detector follows simple and robust 
considerations for particle identification, position, and energy measurements in a non-magnetic environment at energies well 
below the pion production threshold. It consists of 28 elementary cells organized around the beam in a $\varphi$-symmetrical 
arrangement and located at 4 different scattering angles between 6$^{\circ}$ and 15$^{\circ}$ (Fig.~\ref{setup}). The 
elementary cell is placed right after a 100~$\mu$m thick mylar foil acting as vacuum exit window, and is composed of two 
layers of thin scintillator strips followed by a cylindrical BGO (Bi$_4$Ge$_3$O$_{12}$) crystal.

The scintillator hodoscope acts all at once as a neutral particle discriminator, a charged particle tagger, and a position detector. The scintillator strip fired by the scattered electron defines the electron scattering angle. The elementary angular volume, {\it i.e.} the angle binning of experimental data, corresponds to the 20~mm high, 4~mm wide, and 1.3~mm thick strip size. Taylor development techniques allow to transport the strip integrated  experimental cross section at the strip center with a very high accuracy, such that the critical parameter is the actual strip location within the experiment reference frame. The ProRad goal is a full mapping of the detector with an accuracy better than 0.1~mm. \newline
Electrons penetrating the central region of the BGO crystal are tagged by the scintillator hodoscope and absorbed by the 
crystal where they leave an energy deposit signal. The same crystal simultaneously measures elastic and M{\o}ller electrons. 
At each scattering angle, the energy difference between elastic and M{\o}ller electrons allow to distinguish between the two 
processes. The accuracy of this separation is an important parameter of the experiment. It is a convoluted effect of each 
single crystal properties and of the knowledge of the radiative tail associated to elastic and M{\o}ller scattering. The 
ProRad collaboration aims at a better than 0.1\% description of these effects which contribute to the systematics of the 
measurement. 

\begin{figure*}[t!]
\begin{center}
\includegraphics[width=0.995\columnwidth]{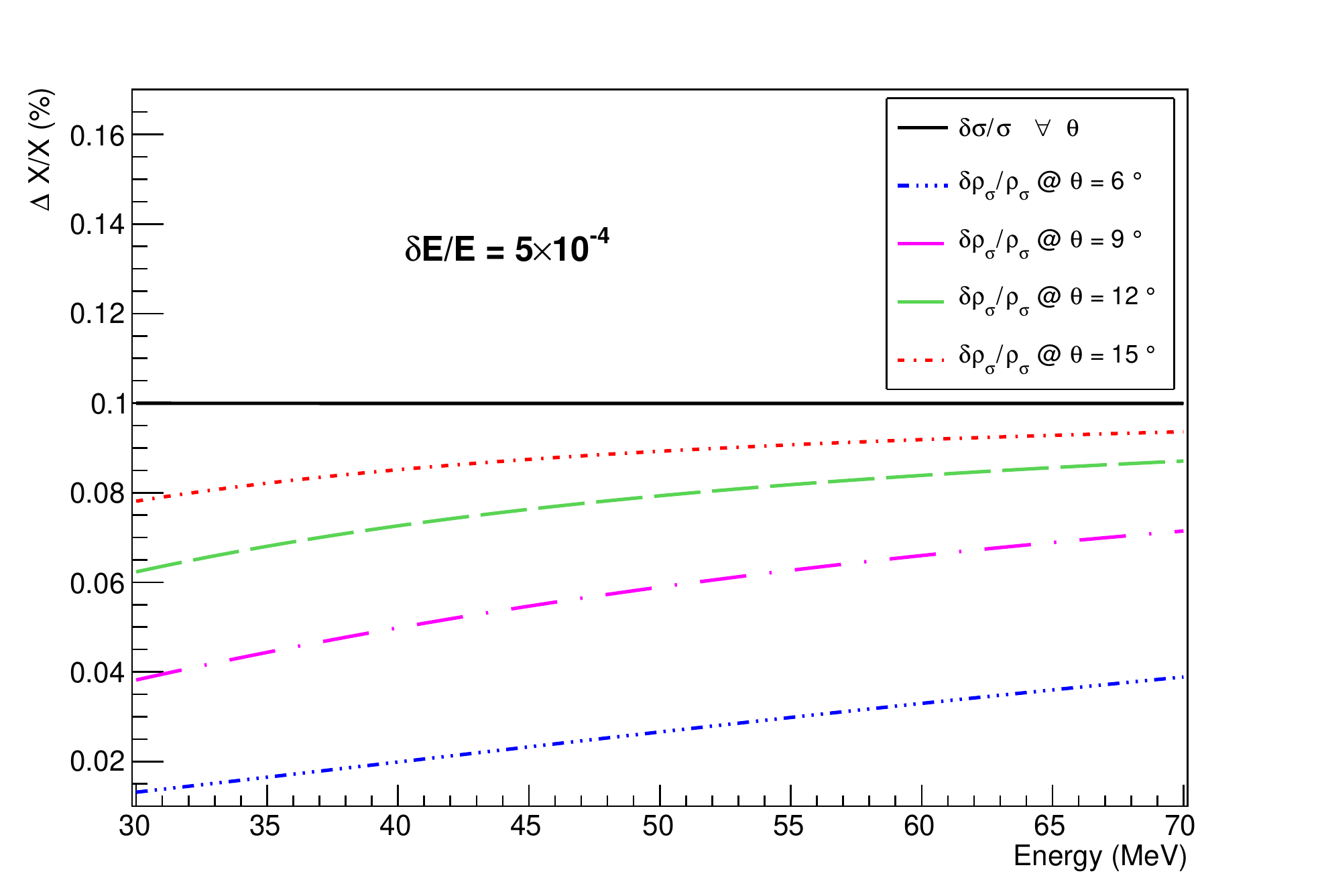}
\includegraphics[width=0.995\columnwidth]{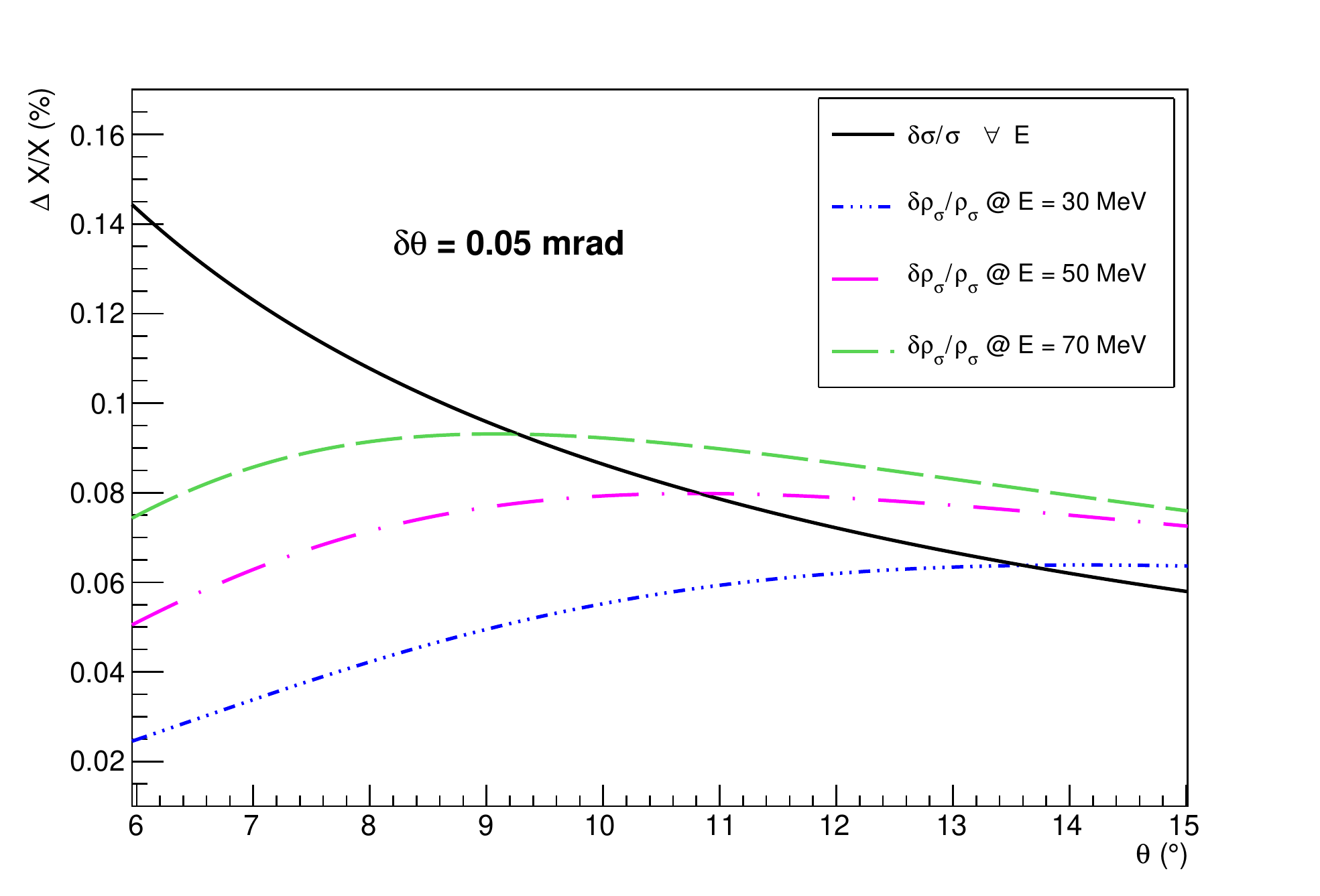}
\caption{Sensitivity of experimental observables (elastic cross section ($\sigma$) or elastic to M{\o}ller cross section ratio ($\rho_{\sigma}$)) to the relative uncertainty on the absolute beam energy (left), and on the uncertainty on the electron scattering angle (right).}
\label{fig:sys}
\end{center}
\end{figure*}

\subsection{Control of systematics}

The electric form factor of the proton is deduced from the experimental cross section ratio according to the expression
\begin{eqnarray}
& & G_E^2(K^2) \nonumber \\
& = & \rho_{\sigma}(K^2) \, \left[ 1 + {\cal O} \left( \frac{K^2}{M^2} \right) \right] \, \frac{\delta_{ee}}{\delta_{ep}} \, { \left( \frac{d^2 \sigma_{ee}}{d\Omega} \bigg{/} \frac{d^2 \sigma_{ep}^{Pt.}}{d\Omega} \right)}_{Th.} \nonumber \\
& - & G_M^2(K^2) \, \left[ {\cal O} \left( \frac{K^2}{M^2} \right) - {\cal O} \left( \frac{m^2}{M^2} \right) \right]
\label{eq:sys}
\end{eqnarray}
where $G_M^2(K^2)$ is the magnetic form factor of the proton and $(m,M)$ are the electron and proton mass. \newline 
In the first term of the right hand side of Eq.~\ref{eq:sys}, the experimental observable ($\rho_{\sigma} (K^2)$) is corrected by several factors: the kinematical coefficient ${\cal O}(K^2/M^2)$; the factor $\delta_{ee}/\delta_{ep}$  originating from radiative effects in $ee$ and $ep$ scatterings; finally, the ratio of the M{\o}ller cross section to the electron elastic scattering cross section off a point-like proton, evaluated theoretically. In the very low momentum transfer region considered at ProRad, the magnetic form factor correction (second term of the right hand side of Eq.~\ref{eq:sys}) is very small and can be evaluated  precisely enough to weakly contribute to the systematics of the measurement. The control of radiative effects with high precision is a challenge that all electron scattering experiments and specifically the low-$K^2$ experimental effort are facing. These have been revisited to take into account lepton mass  effects~\cite{Aku15} of importance at low energies. Further developments to improve the description of higher order  effects~\cite{Arb15,Tom18} are pursued, such that the current theoretical knowledge of radiative effects is reasonably expected to be better than 0.1\%. Additionally, the deconvolution of the energy spectra registered by each detector provides an experimental handle on the radiative tail. \newline
The dominant source of ProRad systematical error originates from the theoretically determined cross section ratio which transfers into the knowledge of the electron beam energy and scattering angle. Measuring the deviation of the beam in a precisely known magnetic field provided by a dipole, the PRAE goal is to determine the absolute beam energy with a 5$\times$10$^{-4}$ accuracy, similarly to the ARC energy measurement~\cite{Mar98} implemented in the Hall A of the Jefferson Laboratory~\cite{Alc04}. Figure~\ref{fig:sys} (left) shows the relative sensitivity of experimental observables in the ProRad energy and angular range at a fixed $\delta E/E$. The expected resolution allows a 0.1\% precision on the elastic cross section, independently of the scattering angle. However, this observable suffers much larger systematics from the target luminosity and the detector efficiency. It is the benefit of the cross section ratio to become independent of these quantities and to further reduce the sensitivity to the beam energy. The knowledge of the angular dispersion of the electron beam, of the location of the interaction  vertex, and of the location of each scintillator hodoscope combine into the electron scattering angle systematics. Thanks to the PRAE beam properties, the quasi-point-like nature of the target, and the precision of the mechanical assembly of the detector, ProRad aims to obtain a 0.05~mrad systematical error on the scattering angle. The effect of this precision on experimental observables is shown on Fig.~\ref{fig:sys} (right) for the beam energies considered at ProRad. It is particularly noticed that the cross section ratio allows to reduce the sensitivity to the angular resolution, and is the only viable observable at small angle to contain angular systematics below 0.1\%.   

%
%
\section{Conclusion}

In summary, this work discusses the correlation between the precision of forthcoming electron scattering experiments at  
low-$K^2$ and their expected impact on the determination of the proton charge radius. It is shown that a 0.1\% precision on 
cross section related experimental observables is necessary to unambiguously establish the value of the proton radius measured 
by the electron scattering technique, {\it i.e.} to obtain a value independent of the maximum momentum transfer ($\vert 
K^2_{max} \vert$) considered for experimental data interpolation in the region limited to 2~fm$^{-2}$. This puts stringent 
metrological constraints on experiments. The technique to meet these requirements in the specific case of the ProRad 
experiment, that will investigate at PRAE the lowest ever measured momentum range, has been presented. \newline
The experimental data fitting procedure has been shown to be an intricate process comprising basic density concerns, the 
region of interpolation, and the fit functional. The convergence criterion of the series representation of the form factor can 
introduce a bias in the density which is minimized if the interpolation region is restricted to 2~fm$^{-2}$. Larger momenta 
region can still be considered within a Fourier transform approach which leads to $k$-dependent functionals instead of the 
commonly used $k^2$-dependence. Within a polynomial approach, it is shown that the polynomial order is constrained to a 
minimal value depending on $\vert K^2_{max} \vert$, and a maximum value constrained by the reliable reconstruction of higher 
order moments of the density. It is most likely a general feature that, with respect to the determination of the proton 
radius, a fit functional is limited to a momentum range of validity. This stresses the importance of pseudo-data methods in 
this problem.

%
%
\begin{acknowledgement}

We would like to thank C\'edric Lorc\'e for enlightening discussion. This work was supported by the LabEx Physique des 2 Infinis et des Origines (ANR-10-LABX-0038) in the framework {\guillemotleft~Investissements d'Avenir~\guillemotright} (ANR-11-IDEX-01), the french Ile-de-France region within the SESAME framework, and the LIA IDEATE. 

\end{acknowledgement}

%
%

%
%
\end{document}